\documentclass[prl,showkeys,twocolumn]{revtex4}
\usepackage{graphicx,amssymb,amsmath}

\begin{document}

\title{A typology of street patterns}

\author{R{\'e}mi Louf}
\email{remi.louf@cea.fr}
\affiliation{Institut de Physique Th\'{e}orique, CEA, CNRS-URA 2306, F-91191, 
Gif-sur-Yvette, France}

\author{Marc Barthelemy}
\email{marc.barthelemy@cea.fr}
\affiliation{Institut de Physique Th\'{e}orique, CEA, CNRS-URA 2306, F-91191, 
Gif-sur-Yvette, France}
\affiliation{Centre d'Analyse et de Math\'ematique Sociales, (CNRS/EHESS) 190-198, avenue de France, 75244 Paris Cedex 13, France}

\begin{abstract}

We propose a quantitative method to classify cities according to their street pattern. We use the conditional probability distribution of shape factor of blocks with a given area, and define what could constitute the `fingerprint' of a city. Using a simple hierarchical clustering method, these fingerprints can then serve as a basis for a typology of cities. We apply this method to a set of $131$ cities in the world, and at an intermediate level of the dendrogram, we observe $4$ large families of cities characterized by different abundances of blocks of a certain area and shape. At a lower level of the classification, we find that most European cities and American cities in our sample fall in their own sub-category, highlighting quantitatively the differences between the typical layouts of cities in both regions. We also show with the example of New York and its different Boroughs, that the fingerprint of a city can be seen as the sum of the ones characterising the different neighbourhoods inside a city. This method provides a quantitative comparison of urban street patterns, which could be helpful for a better understanding of the causes and mechanisms behind their distinct shapes.

\end{abstract}


\maketitle





\section{Introduction}

The recent availability of large amounts of data about urban systems has opened the exciting possibility of a new `Science of Cities', with the aim of understanding and modeling phenomena taking place in the City \cite{Batty:2013}. Urban morphology and morphogenesis, activity and residence location choice, urban sprawl and the evolution of urban networks, are just a few of the important processes that have been discussed for a long time but that we now hope to understand quantitatively. An important component of cities is their street and road network. These networks can be thought as a simplified schematic view of cities, which captures a large part of their structure and organization \cite{Southworth} and contain a large amount of information about underlying and universal mechanisms at play in their formation and evolution. Extracting common patterns between cities is a way towards the identification of these underlying mechanisms. At stake is the question of the processes behind the so-called `organic' patterns---which grow in response to local constraints--- and whether they are preferable to the planned patterns which are designed under large scale constraints. This program is not new \cite{Haggett:1969,Xie:2011}, but the recent dramatic increase  of data availability such as digitized maps, historical or contemporary~\cite{OSM,Strano:2012,Barthelemy:2013,Porta:2014} allows now to test ideas and models on large scale cross-sectional and historical data.

Streets and roads form a network (where nodes are intersections and links are segment roads) which is planar to a good approximation. This network is now fairly well characterized \cite{Jiang:2004,Roswall:2005,Porta:2006,Porta:2006b,Lammer:2006,Crucitti:2006,Cardillo:2006,Xie:2007,Jiang:2007,Masucci:2009,Chan:2011,Courtat:2011}; due to spatial constraints, the degree distribution is peaked, the clustering coefficient and assortativity are large, and most of the interesting information lies in the spatial distribution of betweeenness centrality \cite{Barthelemy:2011}. An important point is that information about these networks is not contained in their adjacency matrix only. Geometry, encoded in the spatial distribution of nodes, plays a crucial role. A classification of cities according to their street network should then rely on both topology and geometry. 

We note that while classifications do not provide any understanding of the objects being classified \emph{per se}, they provide a useful first insight in the different characteristics exhibited by objects of the same nature. Classifying, from a fundamental point of view, is however difficult: finding a typology of street patterns essentially amounts to classifying planar graphs, a non trivial problem. The classification of street networks has been previously addressed by the space syntax community \cite{Hillier,Penn} and a good account can be found in the book by Marshall~\cite{Marshall:2006}. These works, although based on empirical observations, contain a large part of subjectivity and our goal is to eliminate this subjective part to reach a non ambiguous, scientific classification of these patterns. 

An interesting direction is provided by the study of leaves and their classification according to their veination patterns \cite{Katifori:2012,Weitz:2012}, but with a notable difference which prevents us from a direct application to streets: the existence of a hierarchy of veins governed by their diameter (the width of street is usually absent from datasets). Another enticing idea can be found in the mathematics literature: there exists an exact bijection between planar graphs and trees \cite{BDG}. Using this bijection, classifying planar graphs would amount to classify trees, which is a simpler problem. However, this bijection does not take into account the geometrical shape of the planar graph: indeed two street patterns can have the same topology but cells could be of very different areas, leading to patterns visually different and to cities of different structure. It is thus important to take into account not only the topology of the planar graph --- as described by the adjacency matrix --- but also the position of the nodes. In order to do that, we propose in this article a method to characterize this complex object by extracting the `fingerprint' of a street pattern. These fingerprints allow us to define a measure of the distance between two graphs and to construct a classification of cities.

\section{Streets versus blocks}

\begin{figure}[!h]
 \centering
 \includegraphics[width=0.49\textwidth]{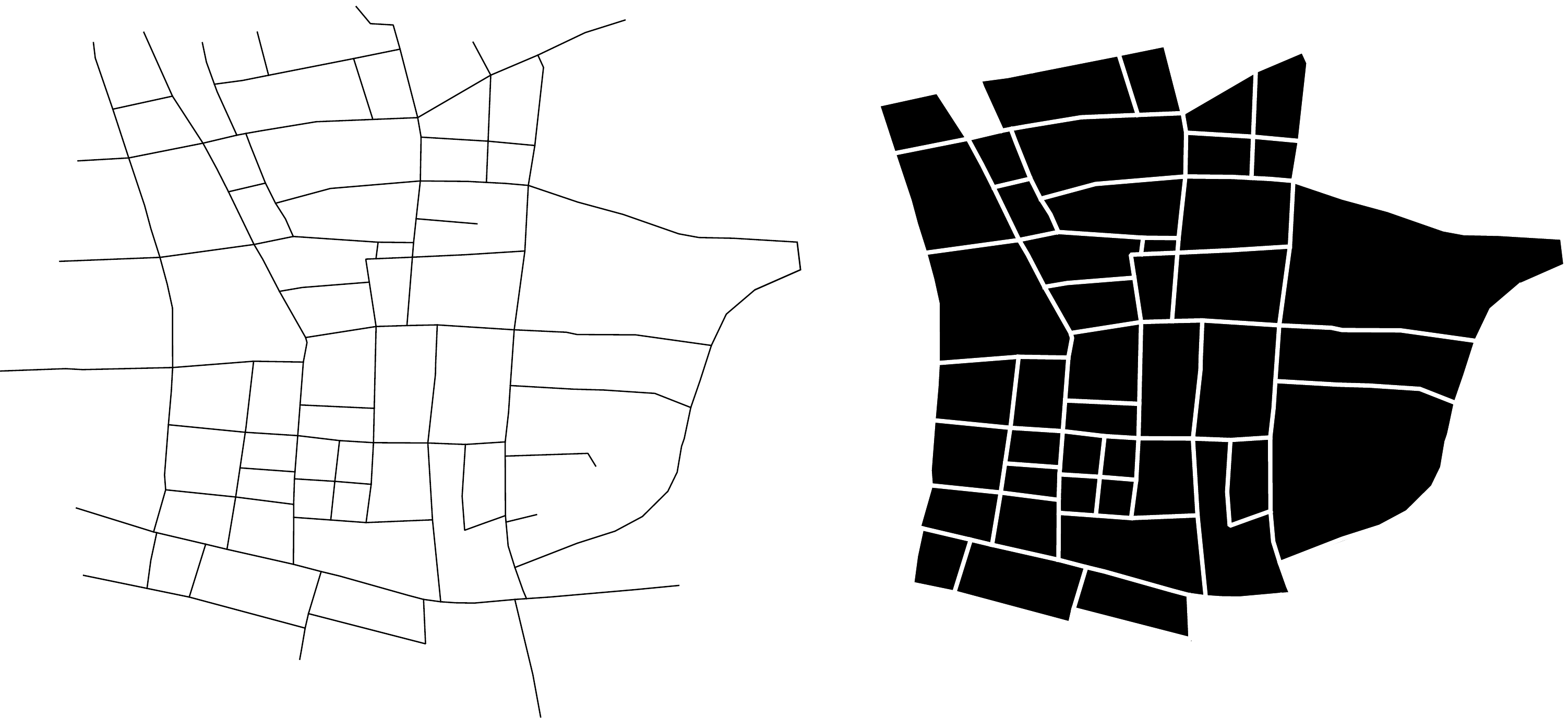}
\caption{{\bf From the street network to blocks.} Example of a street pattern taken in the neighbourhood of Shibuya in Tokyo (Japan) and the corresponding set of blocks. Note that the block representation does not take dead-ends into account.
\label{fig:example}}
\end{figure}

A major shortcoming of existing classifications is that they are based on the street network. This is problematic for two different reasons. First, there is no unambiguous, purely geometrical definition of what a street is: we could define it as the road segment between two intersections, as an almost straight line (up to a certain angular tolerance, see \cite{Con}), or we could also follow the actual street names. There is a certain degree of arbitrariness in each of these definitions, and it is not clear how robust a classification based on streets would be. Second, it seems that what is perceived by the human eye of a city map is not coming from streets but from the distribution of the shape, area and disposition of blocks (see Fig.~\ref{fig:example}). 

A natural idea when trying to classify cities is thus to focus on blocks (or cells, or faces) rather than streets. A block can usually be defined without ambiguity as being the smallest area delimited by roads (it has then to be distinguished from a parcel which is a tax related definition). While the information contained in the blocks and the streets are equivalent (up to dead-ends), the information related to the visual aspect of the street network seems to be easier to extract from blocks. Blocks are indeed simple geometrical objects --- polygons --- whose properties are easily measured. The properties of blocks and their arrangement thus seem to be a good starting point for attempting a classification of urban street patterns.

\section{Characterizing blocks}

\begin{figure*}
 \includegraphics[width=\textwidth]{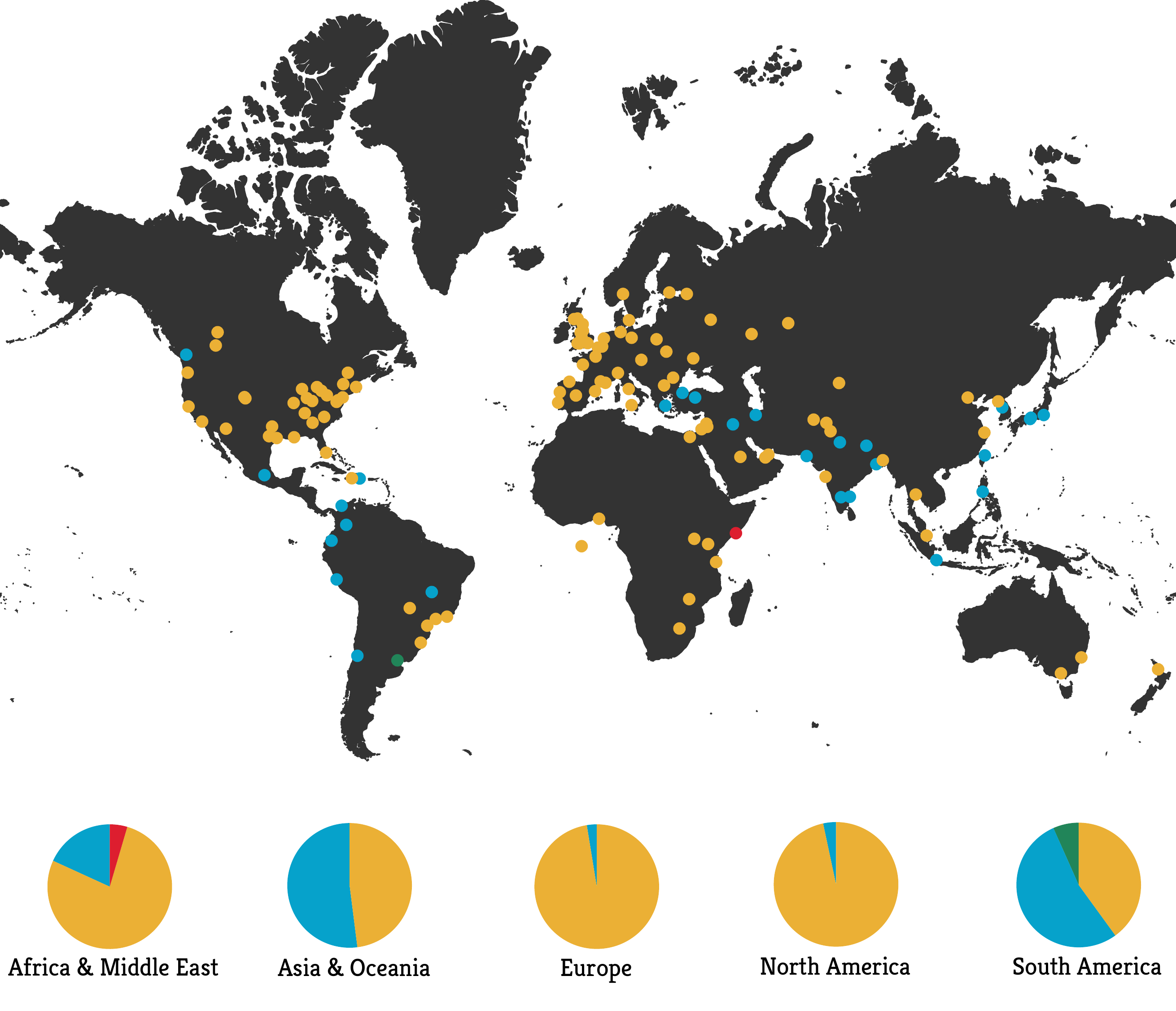}
\caption{{\bf Location of the cities in our dataset and geographical repartition of the different groups.} The color of the dots indicates in which group the city falls, as defined on Fig~\ref{fig:groups}. On the bottom of the map, the pie charts display the relative importance of the different groups per continent for cities in our dataset (Group 1: 0.8\%, Group 2: 20.9\%, Group 3: 77.5\%, Group 4: 0.8\%). We see that the group $3$, composed of cities with blocks of various shapes and a slight predominance of larger areas is by far the most represented group in the world. \label{fig:world_map}}
\end{figure*}

Blocks are defined as the cells of the planar graph formed by streets, and it is relatively easy to extract them from a map. We have gathered road networks for $131$ major cities accross the world, spanning all continents (but Antartica), and their locations are represented on the map Fig.~\ref{fig:world_map}. The street networks have been obtained from the OpenStreetMap database \cite{OSM,Mapzen}, and restricted to the city center using the Global Administrative Areas database (or databases provided by the countries administration). We extracted the blocks from the street network and removed undesired features (faces that have no real-world counterpart but appear due to the particular way data are encoded in OpenStreetMap). We end up with a set of blocks, each with a geographical position corresponding to their centroids. 

Blocks are polygons and as such can be characterized by simple measures. First, the surface area $A$ of a block gives a useful indication, and its distribution is an important information about the block pattern. As in~\cite{Lammer:2006,Fialkowski:2008}, we find that for different cities the distributions have different shapes for small areas, but display fat tails decreasing as a power law
\begin{equation}
P(A)\sim \frac{1}{A^\tau}
\end{equation}
with an exponent of order $\tau\approx 2$ \cite{Lammer:2006,Barthelemy:2011,Strano:2012,Barthelemy:2013}. Although this seemingly universal behaviour gives a useful constraint on any model that attempts at modeling the evolution of cities' road networks, it does not allow to distinguish cities from one another.

\begin{figure*}
\center
\includegraphics[width=\textwidth]{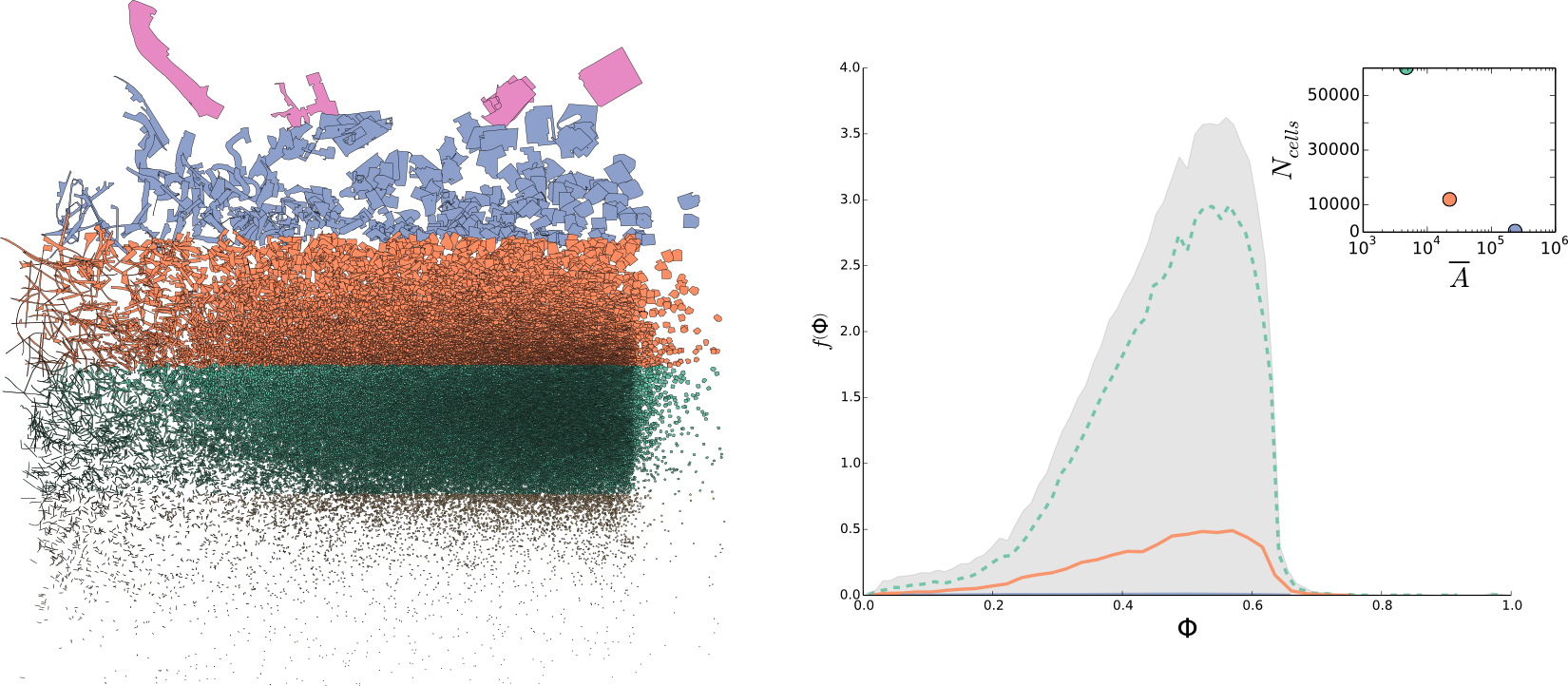}
\includegraphics[width=\textwidth]{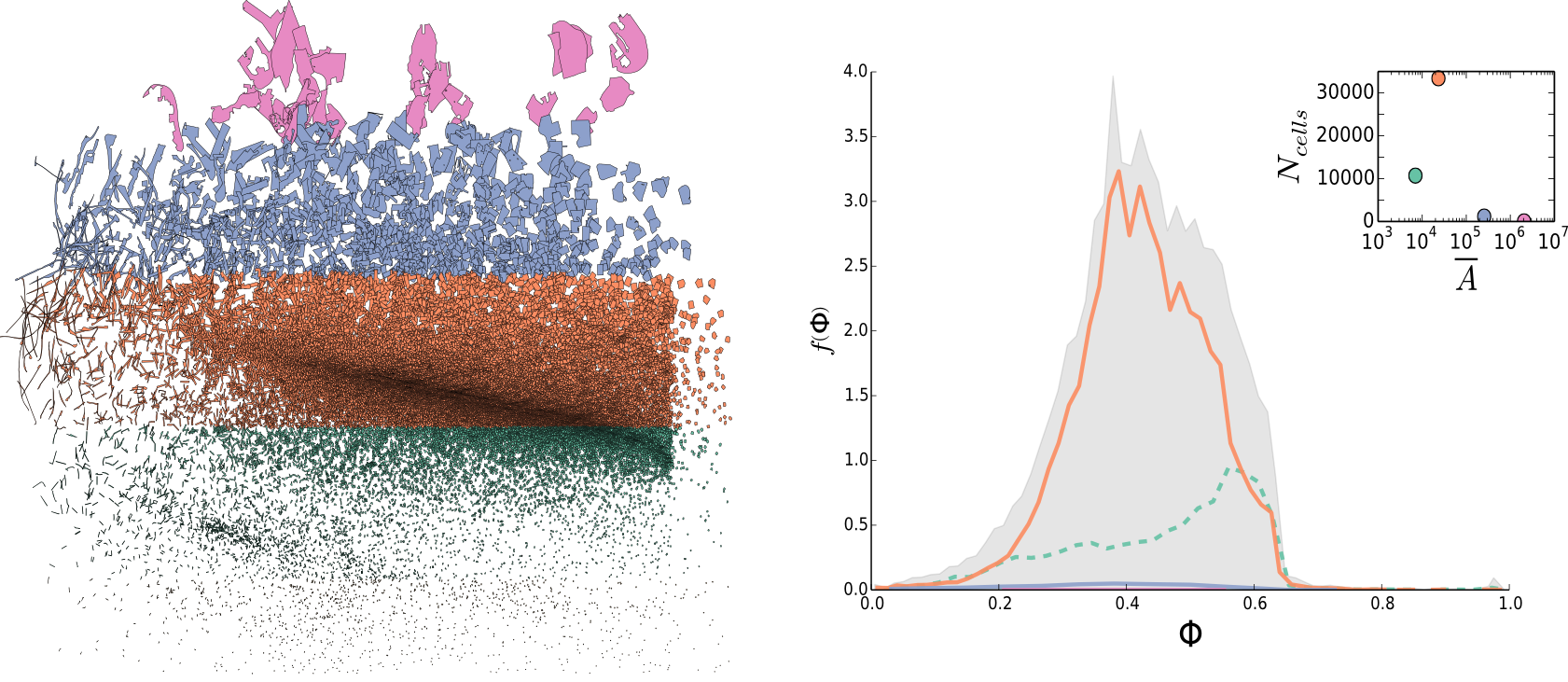}
\caption{{\bf The fingerprints of Tokyo (top) and New-York, NY (bottom).} (Left) We rearrange the blocks of a city according to their area (y-axis), and their  $\Phi$ value (x-axis). The color of each block corresponds to the area category it falls into.  (Right) We quantify this pattern by plotting the repartition of shapes per area category, as measured by $\Phi$ for each area category --- represented by coloured curves. The gray curve is the sum of all the coloured curve and represents the distribution of $\Phi$ for all cells. As shown in the inset, we see that intermediate area categories dominate the total number of cells, and are thus enough for the clustering procedure.\label{fig:fingerprint}}
\end{figure*}

A second characterization of a block is through its shape, with the form (or shape) factor $\Phi$, defined in the Geography literature in~\cite{Hagget:1965} as the ratio between the area of the block and the area of the circumscribed circle $\mathcal{C}$
\begin{equation}
\Phi = \frac{A}{A_{\mathcal{C}}}
\end{equation}
The quantity $\Phi$ is always smaller than one, and the smaller its value, the more anisotropic the block is. There is not a unique correspondence between a particular shape and a value of $\Phi$, but this measure gives a good indication about the block's shape in real-world data, where most blocks are relatively simple polygons. The distributions of $\Phi$ displays important differences from one city to another, and a first naive idea would be to classify cities according to the distribution of block shapes given by $P(\Phi)$. The shape itself is however not enough to account for visual similarities and dissimilarities between street patterns. Indeed, we find for example that for cities such as New-York and Tokyo, even if we observe similar distributions $P(\Phi)$ (see Fig.~\ref{fig:fingerprint}), the visual similarity between both cities's layout is not obvious at all. One reason for this is that blocks can have a similar shape but very different areas: if two cities have blocks of the same shape in the same proportion but with totally different areas, they will look different.  We thus need to combine the information about both the shape and the area.

In order to construct a simple representation of cities which integrates both area and shape, we rearrange the blocks according to their area (on the y-axis) and display their $\Phi$ value on the x-axis (Fig.~ \ref{fig:fingerprint}). We divide the range of areas in (logarithmic) bins and the color of a block represents the area category to which it belongs. We describe quantitatively this pattern by plotting the conditional probability distribution $P(\Phi|A)$ of shapes, given an area bin (Fig.~ \ref{fig:fingerprint}, right). The colored curves represent the distribution of $\Phi$ in each area category, and the curve delimited by the gray area is the sum of all the these curve and is the distribution of $\Phi$ for all cells, which is simply the translation of the well-known formula for probability conditional distribution
\begin{equation}
P(\Phi)=\sum_A\,P(\Phi|A)P(A)
\end{equation}
These figures give a `fingerprint' of the city which encodes information about both the shape and the area of the blocks. In order to quantify the distribution of blocks inside a city, and thus the visual aspect of the latter, we will then use $P(\Phi|A)$ for different area bins. The comparison between these quantities provides the basis for the classification of street patterns that we propose here.

\section{A typology of cities across the world}

Two cities display similar patterns if their blocks have both similar area and shape. In other words, the shape distributions for each area bin should be very close, and this simple idea allows us to propose a distance between street patterns of different cities. More precisely, as one can see on Fig.~\ref{fig:fingerprint}, the number of blocks of area in the range $[10^3,10^5]$ (in square meters) dominate the total number of cells, and we will neglect very small blocks (of area $<10^3\text{m}^2$) and very large ones (of area $>10^5\text{m}^2$). We thus sort the blocks according to their area in two distinct bins
\begin{align*}
\alpha_1 &= \left\{ \text{cells} | \mathcal{A} \in \left[10^3, 10^4\right]\right\}\\
\alpha_2 &= \left\{ \text{cells} | \mathcal{A} \in \left[10^4, 10^5\right]\right\}\\
\end{align*}

\begin{figure*}
 \includegraphics[width=0.75\textwidth]{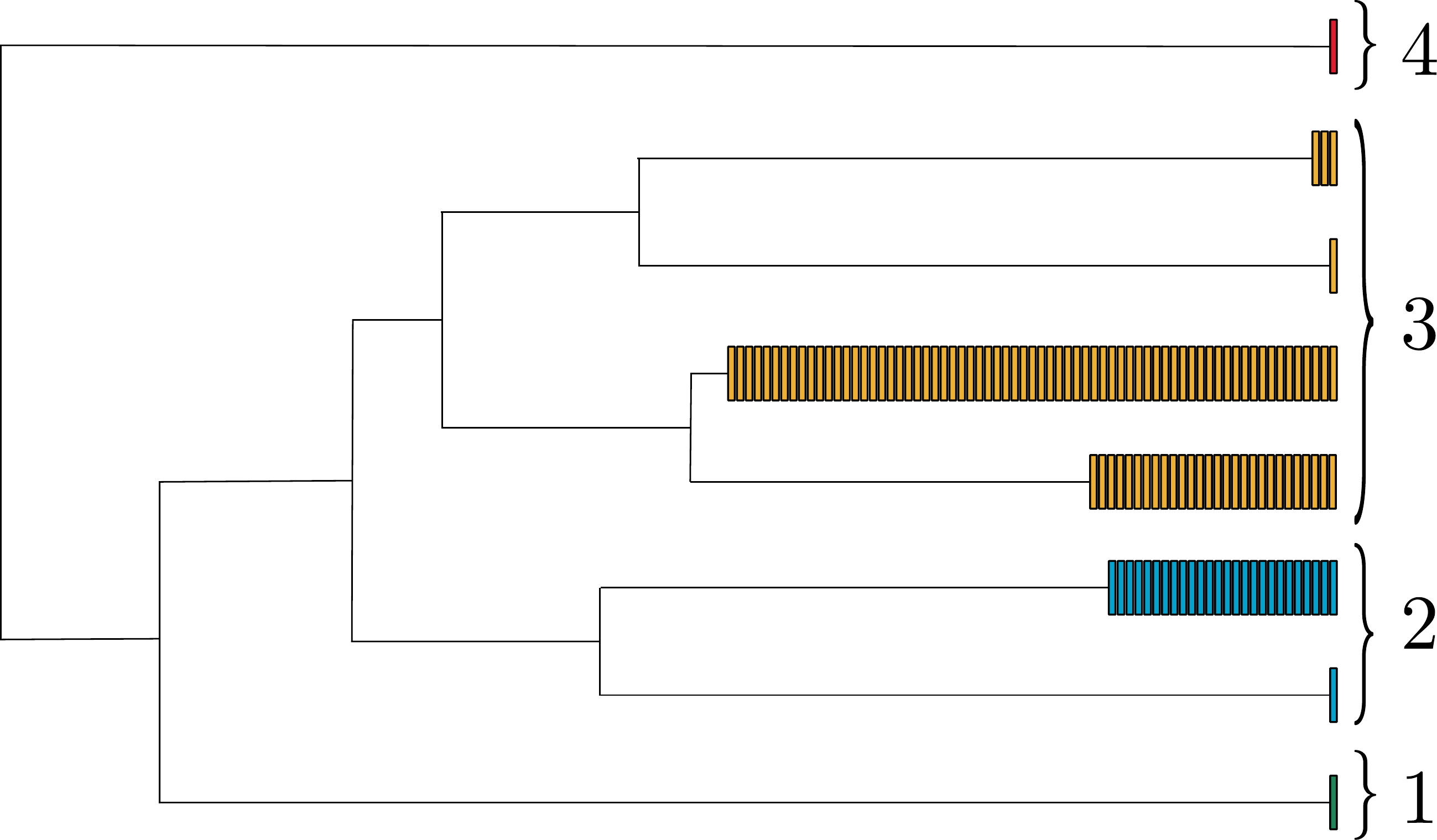}
\caption{{\bf Dendrogram} We represent the structure of the hierarchical clustering at a given level. Interestingly, $68\%$ of american cities are present in the second largest sub-group of group $3$ (fourth from the top). Also, all european cities but Athens are in the largest subgroup of the group $3$ (third from top). This result gives a first quantitative grounding to the feeling that European and most American cities are laid out differently.\label{fig:dendrogram}}
\end{figure*}

\begin{figure*}
\center
 \includegraphics[height=3in]{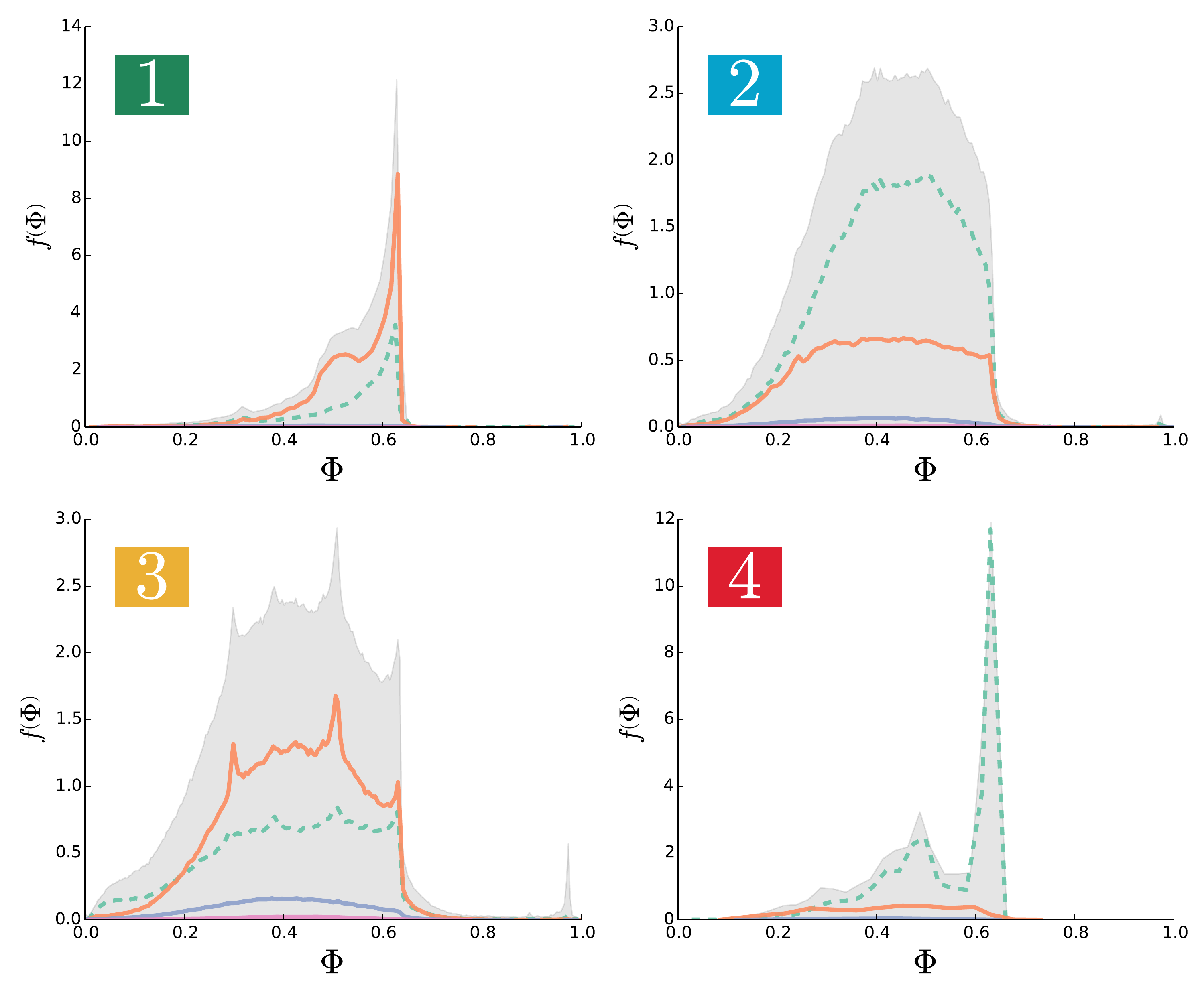}
 \includegraphics[height=3in]{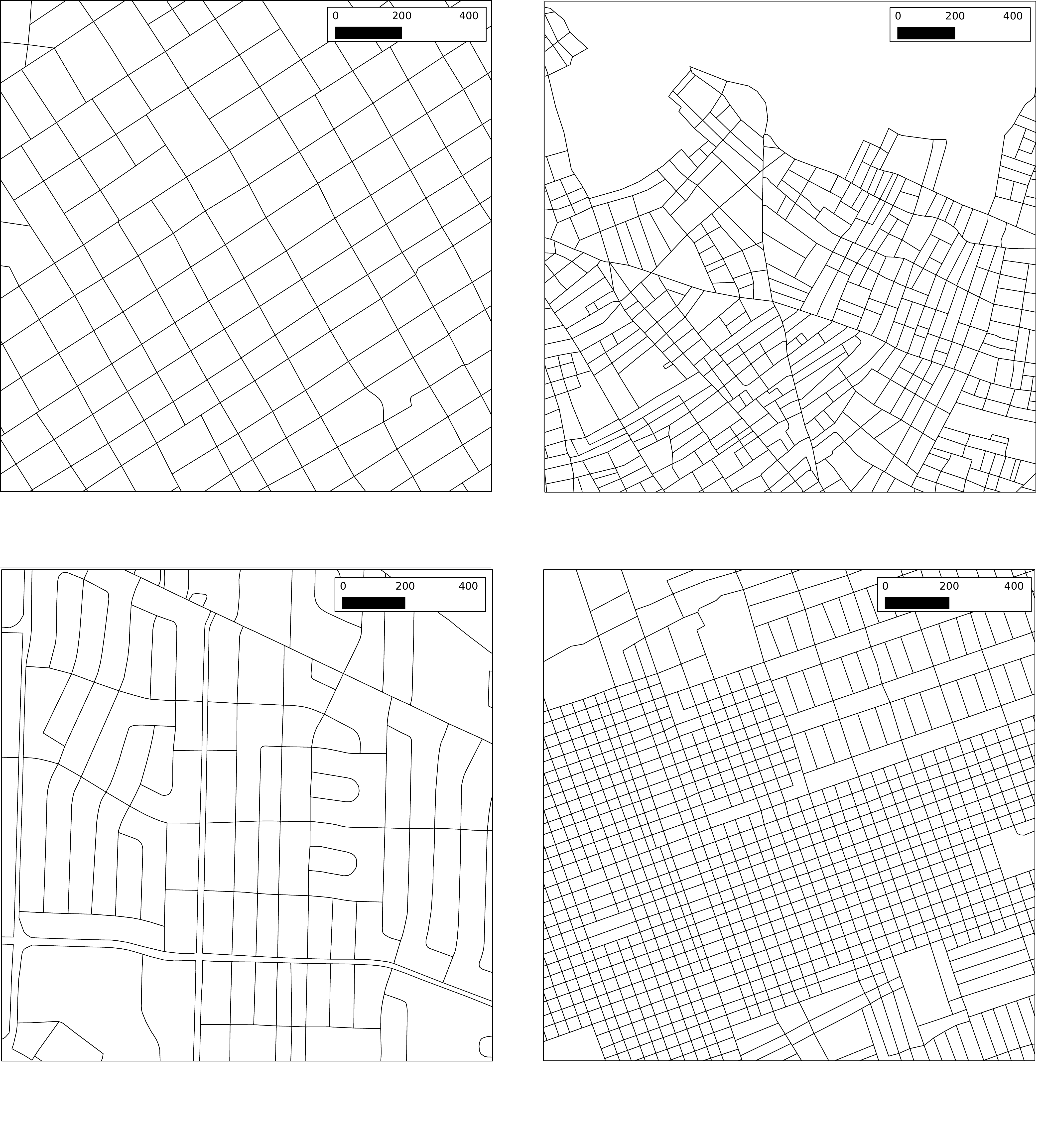}
\caption{{\bf The four groups.} (Left) Average distribution of the shape factor $\Phi$ for each group found by the clustering algorithm (the color of each curve corresponds to the area bin from small to large: dashed green, orange, blue). (Right) Typical street pattern for each group (plotted at the same scale in order to observe differences both in shape and areas). Group 1 (top left): Buenos Aires | Group 2: Athens | Group 3: New Orleans | Group 4: Mogadishu \label{fig:groups}}
\end{figure*}

We denote by $f_\alpha(\Phi)$ the ratio of the number of cells with a form factor $\Phi$ that lie in the bin $\alpha$ over the total number of cells for that city. We then define a distance $d_\alpha$ between two cities $a$ and $b$ characterized by their respective $f^{a}_\alpha$ and $f^{b}_\alpha$
\begin{equation}
d_\alpha(a,b) = \int_0^1\: \left|f^{a}_\alpha(\Phi) - f^{b}_\alpha(\Phi) \right|^n\: \mathrm{d}\Phi
\label{eq:distance}
\end{equation}
We tested different choices ($n=1$ and $n=2$) for $d_\alpha\left(a,b\right)$, and although they might change the position of some cities in the classification, our conclusions are robust. We then construct a global distance $D$ between two cities by combining all area bins $\alpha$
\begin{equation}
D(a,b)= \sum_\alpha d_\alpha(a,b)^{\,2}
\end{equation}
At this point, we have a distance between two cities' pattern and we measure the distance matrix between all the $131$ cities in our dataset, and perform a classical hierarchical clustering on this matrix \cite{Clustering}. We obtain the dendrogram represented on Fig.~\ref{fig:dendrogram} and at an intermediate level, we can identify $4$ distinct categories of cities, which are easily interpretable in terms of the abundance of blocks with a given shape and with small or large area. On Fig.~\ref{fig:groups} we show the average distribution of $\Phi$ for each category and show typical street patterns associated with each of these groups. The main features of each group are the following. 
\begin{itemize}
\item{}
In the group $1$ (comprising Buenos Aires, Argentina only) we essentially have blocks of medium size (in the bin $\alpha_2$) with shapes that are dominated by the square shape and regular rectangles. Small areas (in bin $\alpha_1$) are almost exclusively squares.
\item{}
Athens, Greece is a representative element of group $2$, which comprises cities with a dominant fraction of small blocks with shapes broadly distributed.
\item{}
The group $3$ (illustrated here by New Orleans, USA) is similar to the group 2 in terms of the diversity of shapes but is more balanced in terms of areas, with a slight predominance of medium size blocks. 
\item{}
The group $4$ which contains for this dataset the interesting example of Mogadishu, Somalia displays essentially small, square-shaped blocks, together with a small fraction of small rectangles. 
\end{itemize}

The proportion and location of cities belonging to each group is shown on Fig.~\ref{fig:world_map}. Although one should be wary of sampling bias here, it seems that the type of pattern characteristic of the group $3$ (various shapes with larger areas) largely dominates among cities in the world. Interestingly, all North American cities (except Vancouver, Canada) are part of the group $3$, as well as all European cities (except Athens, Greece). The composition of the other continents is more balanced between the different groups. At a smaller scale within the group $3$ (Fig.~\ref{fig:dendrogram}), all European cities (but Athens) in our sample belong to the same subgroup of the group $3$ (the largest one, third from the top on Fig.~\ref{fig:dendrogram}). Similarly, $15$ American cities out of the $22$ in our dataset belong to the same subgroup of the group $3$ (the second largest one, fourth from the top on Fig.~\ref{fig:dendrogram}. Exceptions are Indianapolis (IN), Portland (OR), Pittsburgh (PA), Cincinnati (OH), Baltimore (MD), Washington (DC), and Boston (MA), which are classified with European cities, confirming the impression that these US cities have a european feel. These results point towards important differences between US and European cities, and could constitute the starting point for the quantitative characterization of these differences \cite{Bretagnolle:2010}.

\section{A local analysis}

\begin{figure*}
 \includegraphics[height=4in]{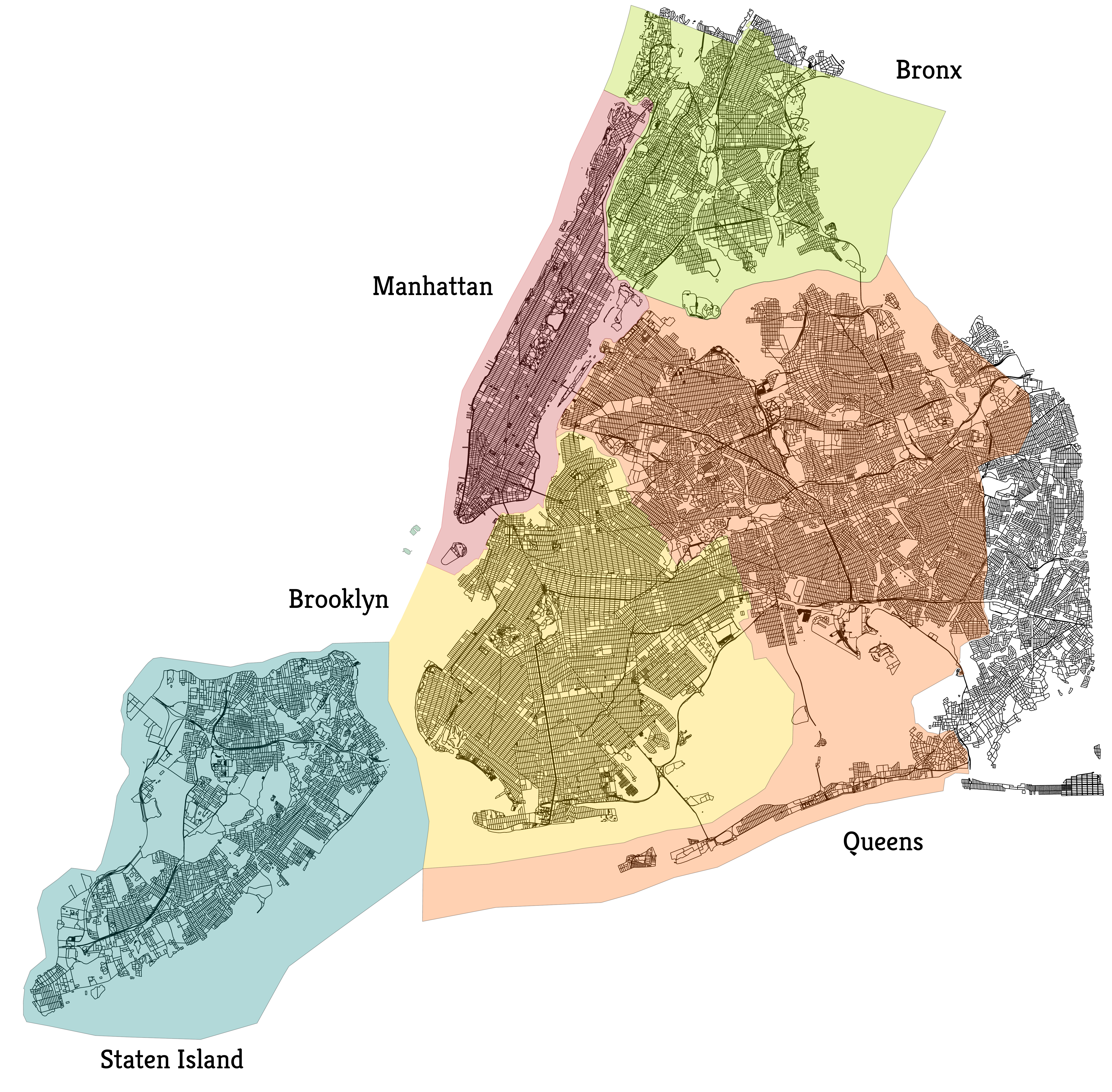}
 \includegraphics[width=\textwidth]{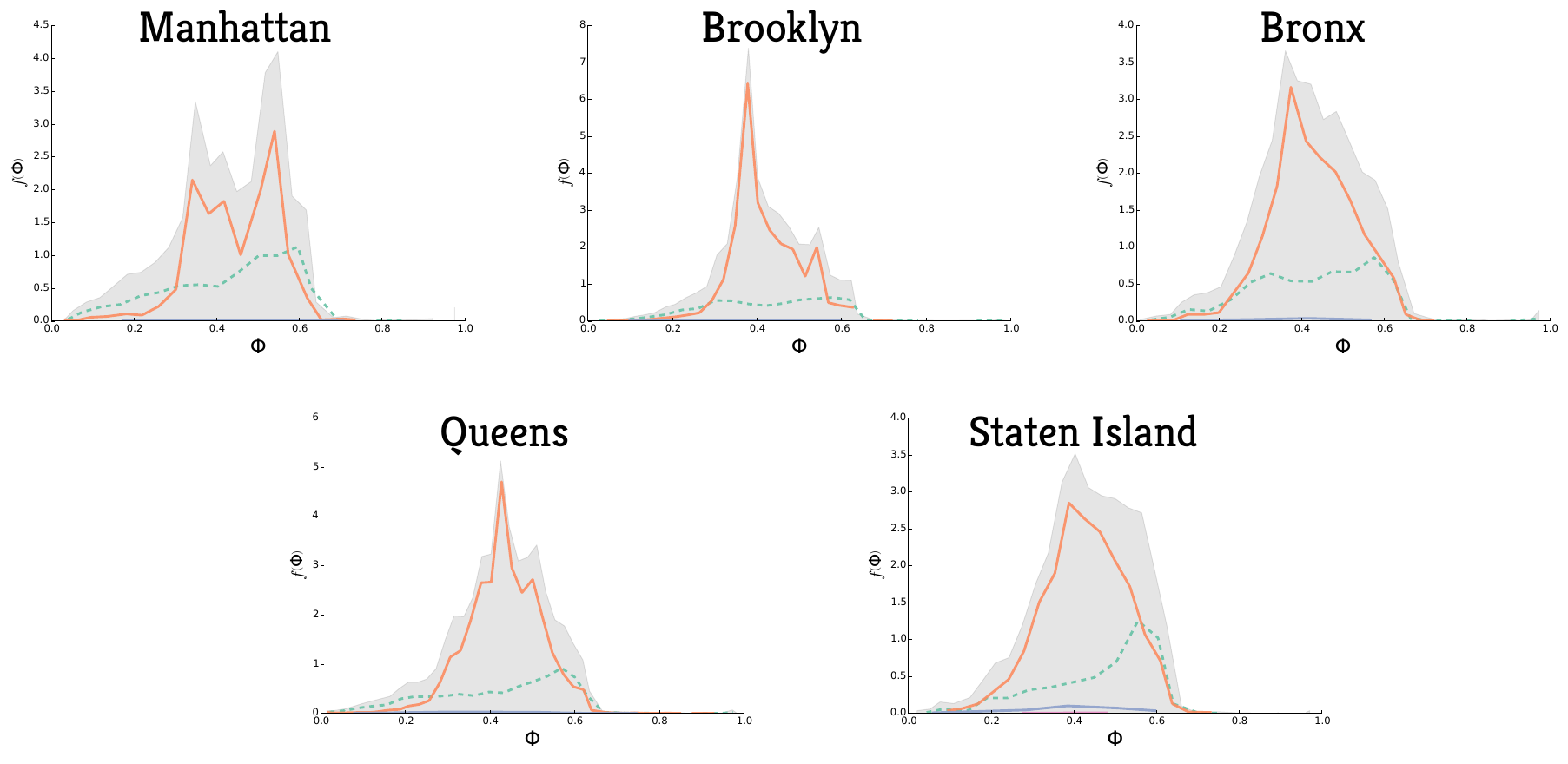}
\caption{{\bf New-York, NY and its different boroughs} \label{fig:ny-boroughs} (Top) We represent New York City and its 5 boroughs: the Bronx, Brooklyn, Manhattan, Queens, and Staten Island. (Bottom) The corresponding fingerprints for each borough. Only Staten Island and the Bronx have similar fingerprints and the others are different. In particular,  Manhattan exhibits two sharp peaks at $\Phi \approx 0.3$ and $\Phi \approx 0.5$ which are the signature of a grid-like pattern with the predominance of two types of rectangles. Brooklyn and Queens exhibit a sharp peak at different values of $\Phi$, signalling the presence of grid-like patterns made of different basic rectangles. }
\end{figure*}

Cities are complex objects, and it is unlikely that a representation as simple as the fingerprint can capture all their intricacies. Indeed, cities are usually made of different neighbourhood which often exhibit different street patterns. In Europe, the division is usually clear between the historical center and the more recent surburbs (a striking example of such differences is the Eixample neighbourhood in Barcelona, very distinct from other areas of the city). In order to illustrate this difference, and to show that they also can be captured with our method, we isolate the different Boroughs of New York, NY: the Bronx, Brooklyn, Manhattan, Queens and Staten Island. We extract the fingerprint of each Borough, as represented on Fig.~\ref{fig:ny-boroughs}. The fingerprint of New York, NY (bottom Fig.~\ref{fig:fingerprint}) is indeed the combination of different fingerprints for each of the boroughs. While Staten Island and the Bronx have very similar fingerprints, the others are different. Manhattan exhibits two sharp peaks at $\Phi \approx 0.3$ and $\Phi \approx 0.5$ which are the signature of a grid-like pattern with the predominance of two types of rectangles. Brooklyn and the Queens exhibit a sharp peak at different values of $\Phi$, also the signature of grid-like patterns with different rectangles for basic shapes. 

\section{Discussion and perspectives}

We have introduced a new way of representing cities's road network that can be seen as the equivalent of fingerprints for cities. It seems reasonable to think that the possibility of a classification based on these fingerprints hints at common causes behind the shape of the networks of cities in the same categories. Of course, the present study has limitations: even if the shape of the blocks alone is good enough for the purpose of giving a rough classification of cities, we miss some aspects of the patterns. Indeed, the way the blocks are arranged together locally should also give some information about the visual aspect of the global pattern. Indeed, many cities are made of neighbourhoods, built at different times, with different street patterns. What is lacking at this point is a systematic, quantitative way to identify and distinguish different neighbourhoods, and to describe the correlation between the blocks's positions. Indeed, the New York Boroughs taken as examples in the last section are administrative, arbitrary definitions of a neighbourhood. Reality is however more complex: similar patterns might span several administrative regions, or a given administrative division might host very distinct neighbourhoods. A further step in the classification would thus be to find a method to extract these neighbourhoods, and integrate the spatial correlations between different types of neighbourhoods.

Despite the simplifications that our method entails, we believe that the classification we propose is an encouraging step towards a quantitative and systematic comparison of the street patterns of different cities. This, together with the specific knowledge of architects, urbanists, etc. should lead to a better understanding of the shape of our cities. Further studies are indeed needed in order to relate the various types that we observe to different urban processes. For example, in some cases, small blocks are obtained through a fragmentation process, and their abundance could be related to the age of the city. A large regularity of cell shapes could be related to planning such as in the case of Manhattan for example, but we also know with the example of Paris \cite{Barthelemy:2013} that a large variety of shapes is also directly related to the effect of a urban modification which does not respect the existing geometry.

\section{Acknowledgements}

We thank Vincenzo Nicosia for interesting discussions at an early stage of this project. We also thank Anne Bretagnolle, Maurizio Gribaudi, Vito Latora, Thomas Louail, Denise Pumain for stimulating discussions at various stages of this study.\\
We would like to thank OSM and all its contributors~\cite{OSM} for making the data we used in this project freely available, as well as the Mapzen mapping lab~\cite{Mapzen} for providing weekly metro extracts.

\section{Data Accessibility}

All the data used in this article can be downloaded from the OpenStreetMap database. Information on how to download OpenStreetMap data are available at http://wiki.openstreetmap.org/wiki/Downloading\_data.

\bibliographystyle{prsty}


\end{document}